\documentclass[%
 reprint,
 amsmath,amssymb,
 aps,
]{revtex4-2}

\usepackage{graphicx}
\usepackage{dcolumn}
\usepackage{bm}

\usepackage{caption}
\usepackage{graphicx}
\usepackage{amssymb}
 \usepackage{amsthm}
\usepackage[dvipsnames]{xcolor}
\usepackage[utf8]{inputenc}
\usepackage{soul}
\usepackage{tabularray}
\usepackage{hyperref}
\usepackage[dvipsnames]{xcolor}
\hypersetup{
    colorlinks,
    linkcolor={green!50!black},
    urlcolor={blue!80!black},
    citecolor   = {red!50!black} 
}

\begin{document}


\title{Effect of strain on the electronic and magnetic properties of bilayer T-phase VS$_2$: A first-principles study}

\author{Mirali Jafari}
 \email{mirali.jafari@amu.edu.pl}
\author{Anna Dyrdał}%
 \email{adyrdal@amu.edu.pl}
\affiliation{Department of Mesoscopic Physics, ISQI, Faculty of Physics, Adam Mickiewicz University, ul. Uniwersytetu Poznańskiego 2, 61-614 Pozna\'n, Poland}

\date{\today}

\begin{abstract}
Using the Density Functional Theory (DFT)  calculations, we determined the electronic and magnetic properties of a T-phase VS$_2$ bilayer as a function of tensile and compressive strain. First, we determine the ground state structural parameters and then the band structure, magnetic anisotropy, exchange parameters, and Curie temperature. Variation of these parameters with the strain is carefully analyzed and described. The easy-plane anisotropy, which is rather small in the absence of strain, becomes remarkably enhanced by tensile strain and reduced almost to zero by compressive strain.
We also show that the exchange parameters and the Curie temperature are remarkably reduced for the compressive strains below roughly -4$\%$.
\begin{description}
\item[Keywords]
$VS_{2}$, Trigonal Phase , DFT , Strain , Electronic and magnetic properties
\end{description}
\end{abstract}

\maketitle

\section{Introduction}
\label{sec:introduction}
Two-dimensional (2D) transition metal dichalcogenides (TMDs) represent a wide class of materials, that have been extensively investigated recently as they are of highly promising potential for applications in nanoelectronic and optoelectronic devices \cite{xiong2017electrostatic, kan2015density, wang2012electronics, mak2010atomically, xiao2012coupled, mak2013tightly, wang2015nonlinear, bertolazzi2011stretching}. Generally, TMDs correspond to a general chemical formula, MX$_{2}$, where M denotes a transition metal element, such as Mo, W, Nb, and V, while X stands for a chalcogen element, like S, Se, and Te. Currently, over 40 different TMDs are known, including metallic, semiconducting, and superconducting ones \cite{marseglia1983transition, wilson1969transition, chhowalla2013chemistry, coleman2011two,jafari2022spin}. Notably, the 2D layered TMDs display distinct physical properties, when compared with their bulk counterparts - especially in the context of band structures \cite {jariwala2014emerging}. Several experimental methods and techniques are currently known, which allow to obtain monolayers of various TMDs, including mechanical exfoliation techniques, liquid exfoliation methods, and chemical vapor deposition (CVD). These techniques collectively facilitate the successful production of various  TMD-monolayers, increasing thus our understanding of these interesting materials \cite{wang2012electronics,radisavljevic2011single, lee2010frictional}.

Typically, layers of vanadium (V) based  TMDs exist in two structural phases, namely the 2H phase characterized by trigonal prismatic coordination, and the 1T phase with octahedral coordination \cite{kan2015density, zhang2013dimension}. Most primitive TMDs lack magnetic properties, which hinders their suitability for applications relying  on magnetism.
Nevertheless, inducing magnetic properties in these materials is feasible with various techniques, such as doping with various point defects, adsorption of non-metal elements, or exploiting edge effects \cite{shidpour2010density, he2010magnetic, wang2011ultra, li2008mos2, ma2011graphene, ma2011electronic}.  For instance, the formation of triple vacancies in single-layer MoS$_{2}$
has been proposed as a tool to generate a net magnetic moment, whereas other defects related to Mo and S atoms do not affect the non-magnetic ground state \cite{xiao2011electrochemically}. In MoS$_{2}$ nanoribbons, the interplay of defects and adsorption can be used for tuning between non-magnetic and magnetic states, depending on the type of defects introduced and the specific sites where the adatoms are adsorbed. However, extending this ability to other TMD materials has proven to be intricate, as the induced magnetic properties are highly dependent on the nature of defects, edge states, and position of dopants, which leads to significant experimental challenges.

On the other hand, computational studies can be used to elucidate the physical properties of TMDs, down to single monolayers. Indeed, such calculations show that monolayers of VX$_{2}$ (where X = S and Se) exhibit intriguing ferromagnetic behavior, providing thus evidence of magnetic properties of pristine 2D monolayers \cite{ma2012evidence}. This insight opens new avenues for the fabrication of ferromagnetic TMDs without resorting to doping with point defects, non-metal element adsorption, or external forces like tensile strain. Encouraged by these theoretical predictions, researchers successfully synthesized ultrathin VS$_{2}$ nanosheets with less than five S–V–S atomic layers, using a modified all-in-solution method. The  corresponding experimental results confirmed the presence of a room temperature ferromagnetism (FM) in the ultrathin VS$_{2}$ nanosheets, accompanied by a very weak antiferromagnetism (AFM). \cite{feng2011metallic,patil2022solvent,Jafari2023}

In our study, we have chosen the T-phase of VS$_{2}$ for detailed investigation, mainly due to its unique and fascinating electronic properties. The T-phase, characterized by octahedral sulfur coordination around vanadium atoms, presents a promising avenue for exploring new phenomena in two-dimensional materials. The T phase is different from its more commonly studied 2H-phase counterpart and exhibits distinctive intrinsic electronic correlations. It is worth noting, that the band structure of the T-phase offers new application possibilities in spintronics and quantum computing \cite{kan2014structures, feng2011metallic}. To get an insight into the fundamental properties of VS$_{2}$, we have decided to investigate the effects of bi-axial strain \cite{zhou2012tensile,jafari2022electronic} on the VS$_{2}$ bilayer configuration. This choice is supported by the following arguments. Firstly, the bilayer structures reveal interlayer interactions and display electronic phenomena that are absent in single-layer counterparts \cite{liu2020magnetoelectric, dong2021comparative}. 
Secondly, bilayer structures are experimentally accessible and are suitable for device applications. Their potential to tailor electronic and magnetic properties  makes them interesting for both fundamental and applied research~\cite{duvjir2020lattice,kawakami2021electronic,wu2022bilayer}.
The remaining part of this paper is structured as follows: Section \ref{sec:methodology} provides an overview of the methodology and computational techniques employed. Section \ref{sec:results} elaborates on the findings derived from the computational analysis. Finally, Section \ref{sec:conclusion} encapsulates the concluding remarks.

\section{Computational details}
\label{sec:methodology}
The first-principles calculations were performed using the Density Functional Theory (DFT) and the Quantum ATK code package (version 2021.06-SP2) \cite{smidstrup2017first}. The calculations were based on the Hohenberg-Kohn theorem \cite{hohenberg1964inhomogeneous} and Kohn-Sham \cite{kohn1965self} equations, and utilized the SG15 collection of optimized norm-conserving Vanderbilt (ONCV) pseudopotentials with Ultra Linear Combination of Atomic Orbitals (LCAO-Ultra) basis set \cite{hamann2013optimized}. The exchange-correlation interaction of electrons was described using the Perdew-Burke-Ernzerhof (PBE) generalized-gradient approximation (GGA) \cite{perdew1996generalized}. The calculations were performed with a converged energy cutoff of 500 Ry and the total energy convergence criteria of $10^{-6}$ eV, with higher criteria of $10^{-8}$ eV for magnetic anisotropy energy calculations. The two-dimensional Brillouin zone was sampled using a $\Gamma$-centered Monkhorst–Pack method \cite{monkhorst1976special} with a k-point grid of 25 $\times$ 25 $\times$ 1. All structures were fully optimized until the force on each atom was less than 0.02 eV/Å. To avoid artificial interaction between image layers, vacuum layers of 25 Angstroms were introduced. Additionally, a weak and non-local van der Waals (vdW) interaction was included in bilayer structures of VS$_2$ material to optimize lattice parameters and bond lengths. The dispersion interactions were accounted for using Grimme DFT-D2 semi-empirical corrections \cite{grimme2006semiempirical}. All structures were fully optimized in the presence of Hubbard U parameter (U=2 eV), where GGA+U was employed to consider the electron-electron correlation effect of the localized $3d$ orbitals of Vanadium (V) \cite{dudarev1998electron}.
\section{Results and discussions}
\label{sec:results}
\subsection{Structural properties}
The T-phase vanadium disulfide (VS$_2$) is a van der Waals layered material, in which an individual monolayer consists of a single layer of vanadium atoms sandwiched between two layers of sulfur atoms, as shown schematically in Fig.\ref{Geometry}. The lattice structure of T-phase VS$_2$ is hexagonal, with each vanadium atom being surrounded by six nearest-neighbor sulfur atoms. The lattice constants of the unit cell in the T-phase VS$_2$ are denoted by $a$ and $b$, which are equal in honeycomb structures, i.e., $a=b$.

\begin{figure}
    \centering
    \includegraphics[width=8 cm]{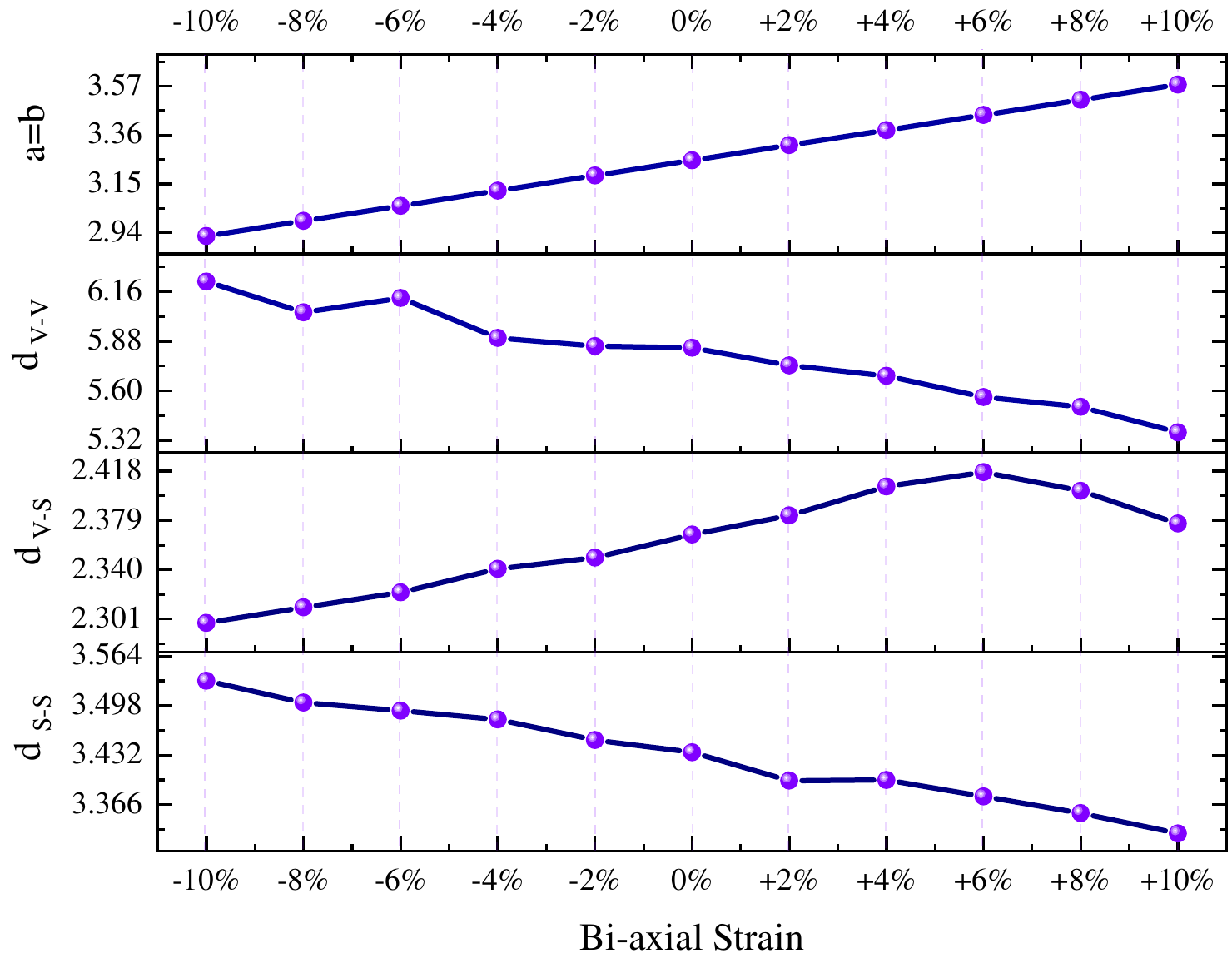}
    \caption{Geometry changes of T-VS$_{2}$ bilayer under the bi-axial strain from -10$\%$ to +10$\%$. Here we show the lattice parameter ($a=b$), the inter-layer distance between Vanadium atoms (d$_{V-V}$), the bond length of Vanadium atoms with the surrounded Sulfur atoms (d$_{V-S}$), and the distance between Sulfur atoms in each layer (d$_{S-S}$). }
    \label{lattice-parameters}
\end{figure}

As illustrated in Fig.\ref{lattice-parameters}, the distance between two Vanadium atoms located in different monolayers in a bilayer, d$_{V-V}$,  exhibits a non-monotonic dependence on the bi-axial strain, increasing smoothly with increasing compressive strain from 0$\%$ to -10$\%$, and decreasing with increasing tensile strain from 0$\%$ to 10$\%$. A weak but noticeable nonmonotonic behaviour appears within the range of -4$\%$ to -8$\%$ of compressive strain, where a relatively fast increase appears in the range of -4$\%$ to -6$\%$,  followed by a  decrease in the range of -6$\%$ to -8$\%$ of compressive strain. This general tendency in the behavior of d$_{V-V}$ with strain can be attributed to the competition between the increasing repulsive forces between the Vanadium atoms under compression and the decreasing attractive forces under tension. The nonmonotonic behavior in the above mentioned range of compressive strains may be due to the occurrence of an energy barrier that needs to be overcome for further compression.

The bond length between Vanadium and Sulfur atoms (d$_{V-S}$) is found to be sensitive to the direction and magnitude of the applied strain. The d$_{V-S}$ exhibits a smooth decrease with increasing compressive strain from 0$\%$ to -10$\%$. Under compressive strain, the compression of the lattice constants enhances the covalent interactions between Vanadium (V) and Sulfur (S) atoms, resulting in a decrease in the bond length between them. Conversely, under tensile strain, the elongation of the lattice constants weakens the covalent interactions between the V and S atoms, resulting in an increase in the bond length between them. However, under tensile strain, d$_{V-S}$ exhibits a non-monotonic variation; it increases for the tensile strain up to 6$\%$, and then smoothly decreases at the higher values of the tensile strains. This anomalous behavior can be explained by the evolution of the electronic structure of the $VS_{2}$ bilayer under strain, which alters the hybridization of the orbitals involved in the V-S bond. Specifically, the tensile strain can induce a weakening of the V-S bond due to the destabilization of the 3d orbital of the Vanadium atom, leading to an initial elongation of d$_{V-S}$. However, at higher tensile strains, the hybridization of the V-S orbitals changes, leading to a stabilization of the 3d orbital of the Vanadium atom and a subsequent contraction of d$_{V-S}$.

The distance between Sulfur atoms, d$_{S-S}$,  is also observed to vary monotonically with increasing compressive and tensile strain, increasing with increasing compressive strain, and decreasing with increasing tensile strain. This trend is due to the changes in the electrostatic interaction between the Sulfur atoms and the Vanadium atoms in the $VS_{2}$ bilayer, which are influenced by the changes in the inter-layer distance and the electronic structure of the bilayer.
\subsection{Static electronic and magnetic properties}
\subsubsection{Electronic bandstructure}
To obtain the spin-resolved electronic band structure, we need to determine first the ground state. 
In the case of the T-phase of VS$_2$, finding the ground state is challenging due to the Coulomb interaction. It has been shown that the ground state of this material can alternate between antiferromagnetic (AFM) and ferromagnetic (FM), when the Coulomb interaction is taken into consideration. This variation in the ground state makes it important to determine where the magnetic moment is predominantly localized. This is especially important when calculating exchange integrals. In this work, we have found that the magnetic moment becomes localized mainly on the vanadium (V) atoms.
To determine the most stable geometry of the T-VS$_2$ bilayer, we calculated the total energy of FM and AFM configurations for different values of $U_{\rm eff}$ (ranging from 0 to 3). We fully optimized the structure without the Coulomb interaction ($U=0$), and found that it had an AFM ground state. Structures with $U_{\rm eff}$ greater than or equal to 1 were found to be FM. We then optimized the structure for  $U_{\rm eff}$ ranging from 1 to 3, and selected the optimized structure with $U_{\rm eff}$=2 eV, based on its stability which was determined by changes in the lattice parameters and bond lengths. Using this value of $U_{\rm eff}$, we calculated all other relevant properties. We employed the DFT calculations to investigate the spin-resolved bandstructure of the bilayer of T-VS$_2$ under bi-axial strain with and without SOC. 
\begin{figure}
    \centering
    \includegraphics[width=8 cm]{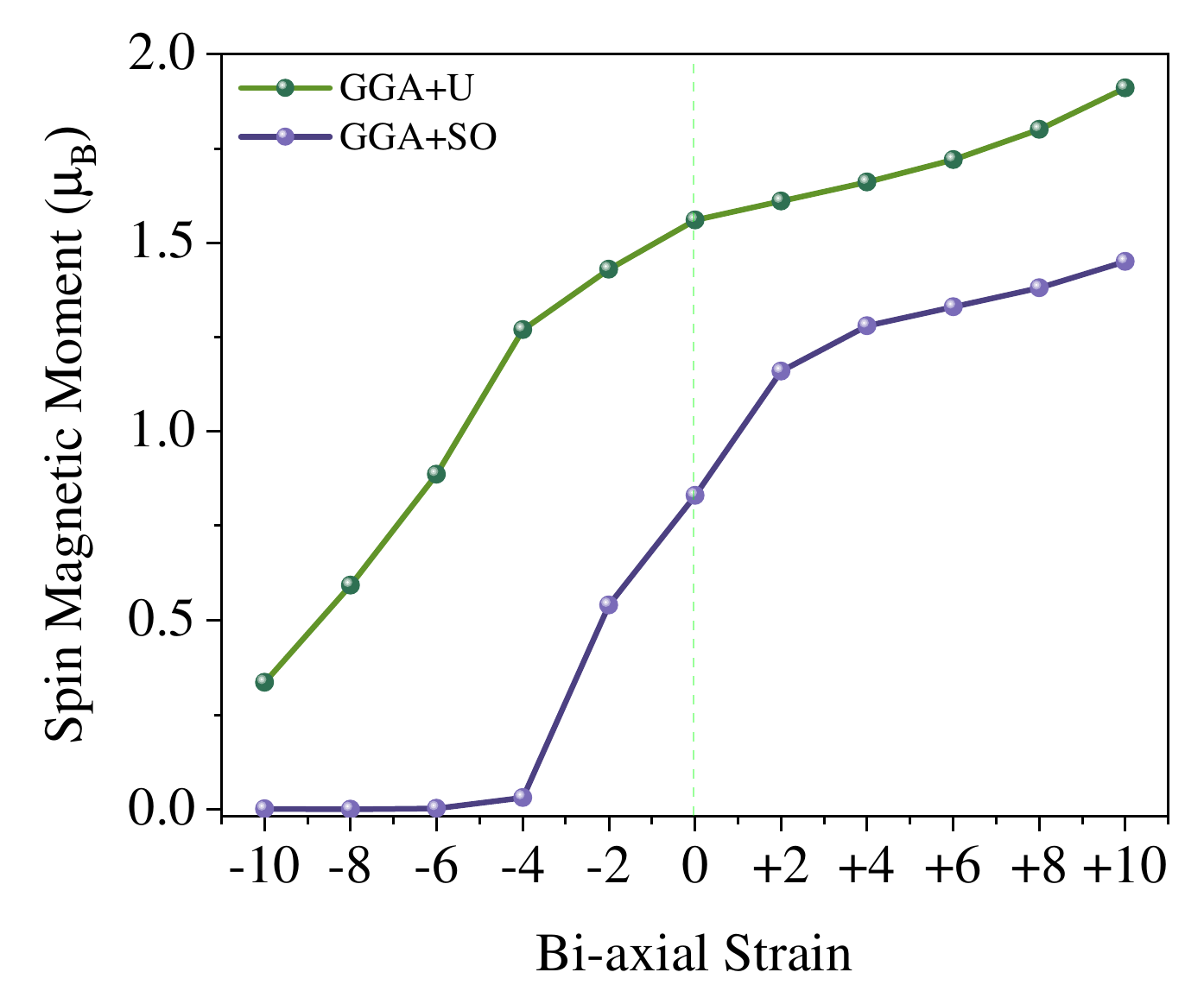}
    \caption{Spin Magnetic Moment (MM) as a function of  applied bi-axial strain }
    \label{MM}
\end{figure}
Our findings, summarized in Table \ref{tab1}, reveal that the unstrained structure exhibits metallic behavior, which is preserved under all compressive strains examined. However, we observe a significant change in the material's behavior at strains of +4$\%$ and +8$\%$, which is attributed to the introduction of SOC. Specifically, the opening of a bandgap at these strains leads to the transition of the bilayer of T-VS$_{2}$ from a metal to a very weak semiconductor. In addition, we computed the spin magnetic moment for each strain using GGA+U and GGA+SOC methods.
Figure \ref{MM} shows the variation of the spin magnetic moment under different strains. Our results demonstrate that the spin magnetic moment increases under the tensile strain, while it decreases under the compressive strain. Moreover, at higher compressive strains, the magnetization of the structure nearly disappears. This behavior could be due to the weakening of the interlayer interaction and the distortion of the crystal lattice. As the lattice compression increases, the magnetic moment decreases until it reaches a critical point where the magnetic order disappears.

\begin{table*}
\centering
\caption{Energy band gap of the structure and spin magnetic moment (MM) under bi-axial strain}
\label{tab1}
\resizebox{0.7\linewidth}{!}{%
\begin{tblr}{
  cells = {c},
  cell{1}{1} = {c=2}{},
  cell{2}{1} = {r=2}{},
  cell{4}{1} = {r=2}{},
  cell{6}{1} = {c=2}{},
  vlines,
  hline{1-2,4,6-7} = {-}{},
  hline{3,5} = {2-13}{},
}
Strain [\%] &  & -10 & -8 & -6 & -4 & -2 & 0 & 2 & 4 & 6 & 8 & 10\\
$E_{Gap}$ [eV] & GGA+U & metal & metal & metal & metal & metal & metal & metal & metal & metal & Half-metal & metal\\
 & GGA+SOC & metal & metal & metal & metal & metal & metal & metal & 0.06 & metal & 0.0004 & metal\\
MM [$\mu_B] $ & GGA+U & 0.336 & 0.593 & 0.887 & 1.27 & 1.43 & 1.56 & 1.61 & 1.66 & 1.72 & 1.80 & 1.91\\
 & GGA+SOC & 0.001 & 0.000 & 0.002 & 0.03 & 0.54 & 0.83 & 1.16 & 1.28 & 1.33 & 1.38 & 1.45\\
MAE [meV] &  & 0 & 0 & 0 & -0.0001 & -0.0335 & -0.0610 & -0.226 & -0.225 & -0.210 & -0.213 & -0.17
\end{tblr}
}
\end{table*}

\subsubsection{Magnetic anisotropy energy}
\begin{figure}
    \centering
    \includegraphics[width=8 cm]{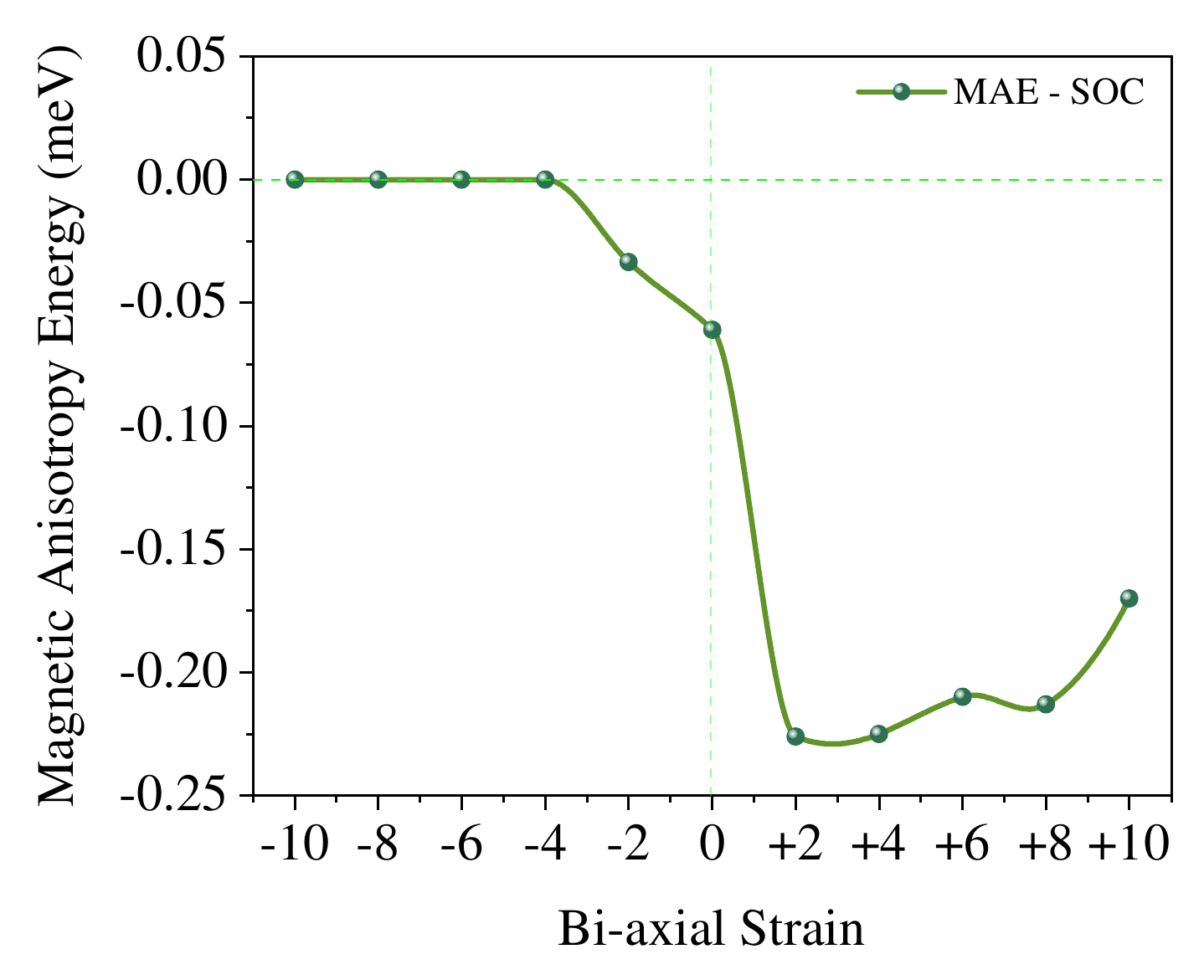}
    \caption{Magnetic Anisotropy Energy changes during the applied bi-axial strain from -10$\%$ to +10$\%$}
    \label{MAE}
\end{figure}
Magnetic anisotropy energy (MAE) plays a crucial role in determining the ground state magnetization orientation and can be calculated using the force theorem by evaluating the energy difference between relevant spin orientations. In this paper, we focus on the perpendicular 
anisotropy, which is defined as the energy difference between magnetizations along two specific crystallographic orientations. More precisely, the perpendicular anisotropy is defined as ${\rm MAE}=E_{[100]}-E_{[001]}$ (or ${\rm MAE}=E_{[010]}-E_{[001]}$, when the system is magnetically isotropic or near isotropic in the plane), where the positive value of ${\rm MAE}$ corresponds to the perpendicular easy axis, while the negative value to the perpendicular hard axis, i.e. to the easy-plane anisotropy. Our results, as illustrated in Figure \ref{MAE}, show that the pure T-VS$_2$ structure without any strain exhibits a relatively small easy-plane magnetic anisotropy, which increases under the tensile strain due to the increase in crystal field splitting. In contrast, under compressive strain, the easy-plane magnetic anisotropy decreases and becomes close to zero, then for the strain below $\leq -4\%$ is roughly equal to zero.
This result can be attributed to the decrease in crystal field splitting, which makes the system more isotropic.

\begin{figure}
    \centering
    \includegraphics[width=8 cm]{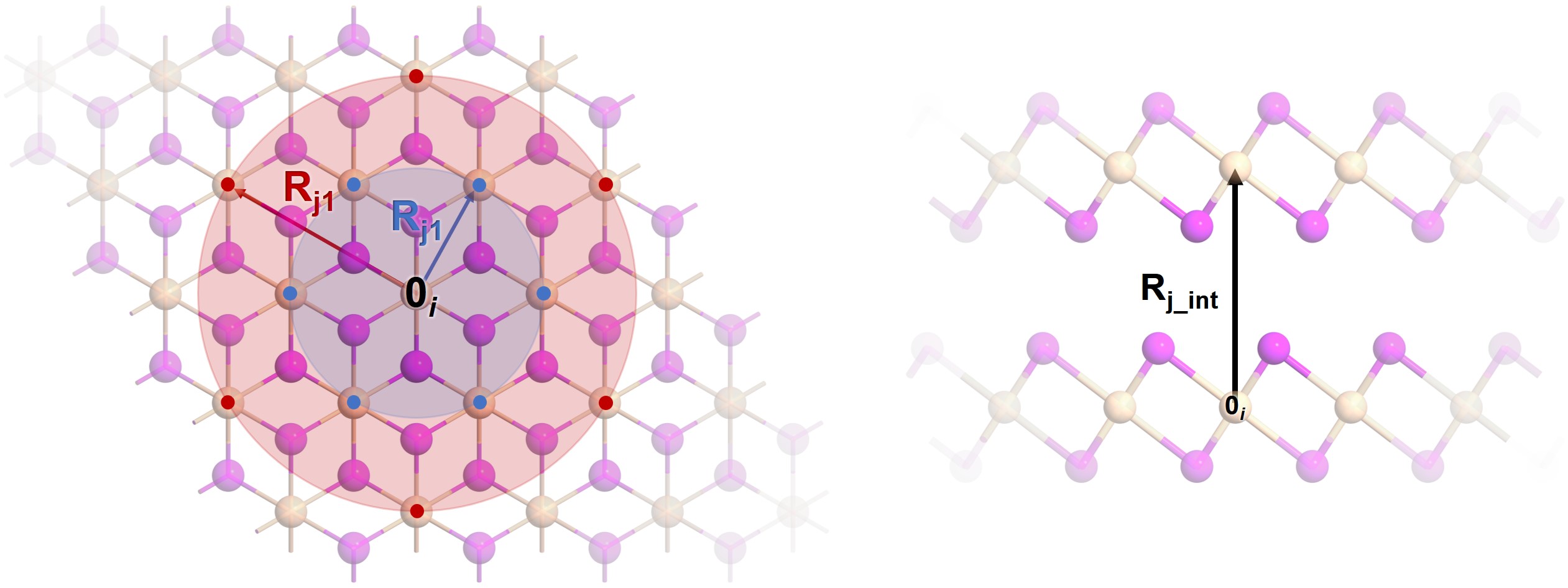}
    \caption{Schematic representation of the nearest ($J_1$, blue) and next-nearest ($J_2$, red) neighbors surrounding the central Vanadium atom (labeled as 0) in the $T-VS_2$ bilayer. The right panel presents  a side view emphasizing interlayer atoms ($J_{int}$.).}
    \label{lkag}
\end{figure}

\subsection{Exchange parameters, Curie temperature}
\begin{table*}[t]
\centering
\caption{Changes in the exchange parameters  J$_1$, J$_2$, and J$_{int}$ under the bi-axial strain}
\label{tab1}
\resizebox{0.66\linewidth}{!}{%
\begin{tblr}{
  cells = {c},
  cell{1}{1} = {c=2}{},
  cell{2}{1} = {r=2}{},
  cell{4}{1} = {r=2}{},
  cell{6}{1} = {r=2}{},
  vlines,
  hline{1-2,4,6,8} = {-}{},
  hline{3,5,7} = {2-13}{},
}
Strain [\%] &  & -10 & -8 & -6 & -4 & -2 & 0 & 2 & 4 & 6 & 8 & 10\\
J$_1$ [meV] & GGA+U & -0.09 & -0.04 & -0.20 & -0.50 & 7.41 & 7.86 & 6.92 & 6.65 & 6.34 & 5.91 & 5.38\\
 & GGA+SOC & 0 & 0 & 0 & 0 & 0.45 & 1.51 & 7.15 & 7.17 & 6.12 & 5.62 & 5.70\\
J$_2$ [meV] & GGA+U & 0.40 & 1.05 & 1.23 & -0.46 & -0.28 & 1.66 & 0.52 & 2.32 & 2.40 & 2.38 & 2.29\\
 & GGA+SOC & 0 & 0 & 0 & 0 & 0.92 & 0.93 & 0.26 & 1.61 & 1.94 & 2.04 & 2.11\\
J$_{int}$ [meV]  & GGA+U & 0.02 & -0.26 & -0.01 & 0.17 & 0.96 & 0.77 & 2.14 & 0.53 & 0.30 & 0.08 & -0.17\\
 & GGA+SOC & 0 & 0 & 0 & 0 & 0.09 & 0.24 & 0.35 & 0.35 & 0.26 & 0.11 & -0.07
\end{tblr}
}
\end{table*}

\begin{figure*}[htp]
    \centering
    \includegraphics[width=2\columnwidth ]{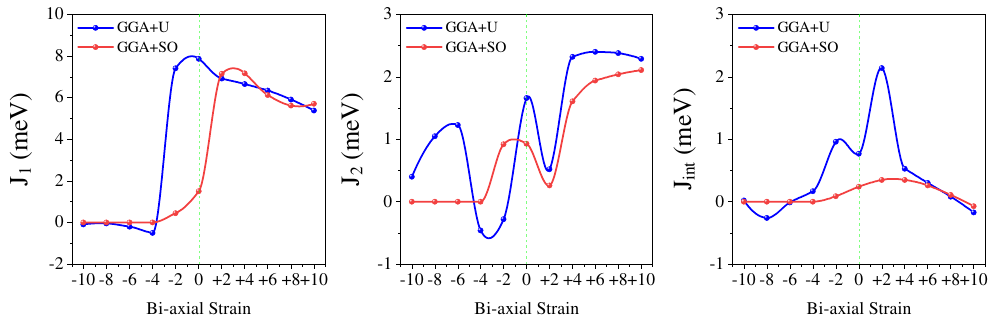}
    \caption{Exchange coupling constants as a function of applied bi-axial strain from -10$\%$ to +10$\%$}
    \label{Jij}
\end{figure*}

When investigating magnetic properties of materials, the density functional theory (DFT) technique is commonly used as it gives reliable results. This technique allows  to determine material parameters, like zero-temperature magnetization and magnetic ordering.
Additionally, DFT enables calculation of the exchange parameters  
which enter  Heisenberg spin Hamiltonian -- commonly used to determine 
the Curie temperature, spin wave dispersions, and others.
The Heisenberg exchange Hamiltonian can be written as:
\begin{equation}
H = -\sum_{i\neq j} J_{ij}\hat{e}_i\hat{e}_j
\end{equation}
Here, $\hat{e}_i$ represents the normalized local spin vector on atom $i$, and $J_{ij}$ stands for the Heisenberg exchange coupling constant. We utilized the energy mapping technique developed by 
Liechtenstein-Katsnelson-Antropov-Gubanov (LKAG) formula (1987) \cite{liechtenstein1987local}  to determine the Heisenberg intra-layer exchange coupling constants between the nearest-neighbors ($J_1$) and next nearest-neighbors ($J_2$), as well as for the inter-layer coupling ($J_{\rm int}$).
%
To compute the exchange coupling matrix element $J_{0i, Rj}$, which describe the interaction between atom $i$ in the central unit cell (labeled $0$) and atom $j$ in a different unit cell displaced from the central one by a lattice vector $R$, as shown in Figure \ref{lkag}, the real-space Green's function method is employed~\cite{liechtenstein1987local}, as implemented in the  ATK package.



Our results for the clean and unstrained structure show that both  $J_1$ and $J_2$ are positive, indicating that the magnetic interactions between the nearest and next nearest neighbors are ferromagnetic. The parameter $J_{\rm int}$ is also positive, indicating that the magnetic interaction between spins in different layers is ferromagnetic, too. Furthermore, we found that $J_1$ is the largest coupling parameter, 
$J_2$ is smaller than $J_1$, 
and $J_{int}$ is the smallest from the set of the three determined coupling constants.

Our calculations of the Heisenberg exchange coupling constants for the bilayer of T-phase VS$_2$ reveal several interesting behaviors. First, we find that both GGA+U and GGA+SOC methods yield similar behavior for $J_1$, but with small differences in magnitude. Specifically, the $J_1$ values obtained from both methods are almost the same at zero strain, with the GGA+U values being slightly larger than the GGA+SOC values. This can be attributed to the fact that the GGA+U method tends to overestimate the strength of the Coulomb interaction, leading to larger values of the exchange coupling constants. However, at higher tensile strains, the $J_1$ values obtained from GGA+U and GGA+SOC calculations converge, and the difference between them becomes smaller. This behavior can be explained by a combination of strain-induced changes in the electronic structure, spin polarization, and crystal structure of the bilayer. As the magnitude of the tensile strain increases, the crystal lattice becomes more distorted, which leads to a modification of the exchange pathways and a consequent change in the magnetic properties of the system. Additionally, the spin-orbit coupling becomes more significant at higher strains, leading to a more pronounced influence on the magnetic properties of the system. These changes in the electronic structure and spin polarization can lead to $J_1$  obtained from GGA+U and GGA+SOC  more similar at higher tensile strains. In the case of compressive strains, we observed that both GGA+U and GGA+SOC result in a decrease in $J_1$ due to the crystal lattice distortion caused by the compressive strain, which in turn leads to a change in the magnetic properties of the system. Notably, we found that at the strain of -4$\%$, the sign of $J_1$ obtained from GGA+U calculations changes to antiferromagnetic coupling, whereas the value obtained from GGA+SOC approaches zero. This can be explained by the tendency of the GGA+U method to overestimate the Coulomb interaction strength, leading to larger values of the exchange coupling constants and its sign change. In contrast, the GGA+SOC method includes the spin-orbit coupling, which can suppress the exchange interaction and can result in smaller values of the exchange parameters. 
We observed a similar trend for the next nearest neighbor exchange coupling constant ($J_2$), with both GGA+U and GGA+SOC showing fluctuations in their magnitude. From Figure \ref{Jij}, it can be clearly seen that the magnitude of $J_2$ fluctuates due to the changes in the crystal structure under strain. Interestingly, in the presence of Hubbard corrections, we observe the sign change of  $J_2$ from FM to AFM  and the from AFM to FM under compressive strain. 

Finally, we also evaluated the inter-layer exchange coupling constant ($J_{\rm int}$), which determines the ground state of the bilayer structure of T-VS$_2$. From Figure \ref{Jij} follows that the inter-layer exchange coupling constant sharply increases at strains of $\pm 2\%$, indicating the preference of the ferromagnetic coupling in this range of strains. However, for higher tensile and compressive strains, the inter-layer exchange coupling constant decreases, indicating the tendency towards antiferromagnetic coupling. This behavior can be attributed to the interplay between changes in the crystal structure and the electronic and magnetic properties of the system. Interestingly, strains equal to or larger than $-6 \%$ lead to deformation of the crystal structure and a change from a ferromagnetic to an antiferromagnetic state.

Having found the exchange parameters, one can determine the Curie temperature. To do this we use the Mean Field Approximation (MFA) and the Random Phase Approximation (RPA). In the Mean Field Approximation, spins are treated based on an average interaction field, and any correlations beyond the nearest neighbors are disregarded. The MFA allows for a straightforward calculation of the Curie temperature (T$_c$) using the following expression:

\begin{table*}[t]
\centering
\caption{Curie Temperature ($T_c$) obtained by MFA and RPA methods for different applied bi-axial strains.}
\label{tabTc}
\resizebox{0.67\linewidth}{!}{%
\begin{tblr}{
  cells = {c},
  cell{1}{1} = {c=2}{},
  cell{2}{1} = {r=2}{},
  cell{4}{1} = {r=2}{},
  vlines,
  hline{1-2,4,6} = {-}{},
  hline{3,5} = {2-13}{},
}
Strain [\%] &  & -10 & -8 & -6 & -4 & -2 & 0 & 2 & 4 & 6 & 8 & 10\\
MFA & GGA+U & 16.2 & 62.3 & 30.1 & - & 212.4 & 394.2 & 439.5 & 463.0 & 469.8 & 463.5 & 439.1\\
 & GGA+SOC & 0 & 0 & 0 & 0.3 & 66.5 & 92.1 & 276.8 & 410.7 & 420.1 & 424.3 & 442.1\\
RPA & GGA+U & 4.3 & 56.4 & 4.3 & 85.5 & 123.3 & 271.5 & 345.9 & 345.1 & 372.1 & 403.2 & 391.3\\
 & GGA+SOC & 0 & 0 & 0 & 0.3 & 59.2 & 85.0 & 185.4 & 301.0 & 332.5 & 302.5 & 269.6
\end{tblr}
}
\end{table*}

\begin{equation}
T_{c} = \frac{2}{3 k_{B}}
\left( Z_{1} J_{1}+Z_{2}  J_{2}+Z_{int}  J_{int}\right),
\end{equation}
where k$_B$ is the Boltzmann constant, while $Z_{1}=6$, $Z_{2}=6$, and $Z_{int}=1$ denote the respective numbers of  the nearest neighbors, next-nearest neighbors, and interlayer nearest neighbors atoms. In turn, the Random Phase Approximation (RPA) takes into account fluctuations in the spin orientations and incorporates correlations beyond the mean field. This approximation offers a more precise estimation of the Curie temperature. However, the RPA involves solving a system of coupled equations that take into account the spin correlation functions, resulting in a more intricate expression for T$_c$.
\begin{figure}
    \centering
    \includegraphics[width=8 cm]{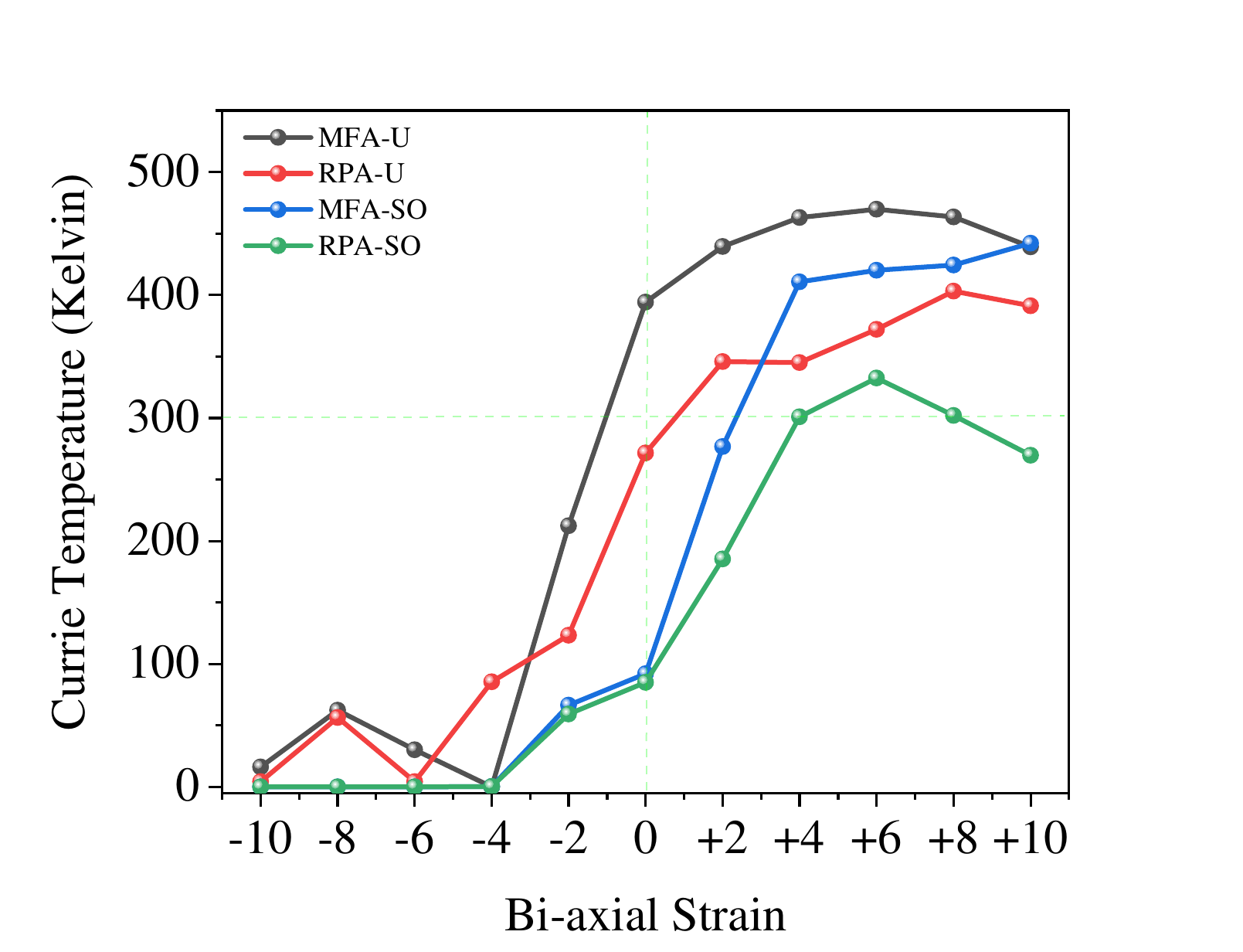}
    \caption{The estimated Curie temperature in the Mean Field Approximation (MFA) and Random Phase Approximation (RPA) under bi-axial strain.}
    \label{TC}
\end{figure}
The pure structure of the bilayer of T-VS$_2$ has $T_{c}$ close to or beyond the room temperature for GGA+U in the MFA and RPA methods, respectively. However, this value is lower for the GGA+SOC calculations. Furthermore, we have calculated the Curie temperature for the different strains and the results are listed in Table \ref{tabTc}. The T$_c$ plot as a function of bi-axial strain exhibits a similar trend to that of the exchange parameters, as shown in Figure \ref{TC}. Under tensile strain, the Curie temperature shows an increasing tendency. However, at higher tensile strains, it slightly decreases, which can be attributed to the interplay between the changes in electronic structure, spin interactions, and lattice distortions.
In turn, the Curie temperature under compressive strains decreases, as anticipated from the corresponding $J_1$ and $J_2$ interactions. The decrease in Curie's temperature continues until certain points where it becomes very low or even vanishes.
\section{Summary and conclusions}
\label{sec:conclusion}
In this paper, we have analyzed the influence of biaxial compressive and tensile strains on the electronic and magnetic properties of a bilayer of  T-VS$_2$. The bilayer structure seems to be interesting for applications, as it reveals the properties of a natural spin valve {\color{blue}\cite{jafari2022spin,elahi2022brief,long2020stacking,wu2021high}}. We have determined the strain-induced changes in the structural parameters, electronic band structure, magnetic anisotropy, exchange parameters, and Curie temperature. 

An interesting effect is that easy-plane magnetic anisotropy is enhanced for tensile strains and reduced almost to zero for compressive strains. This anisotropy in unstrained structures is rather small,  so the tensile strain can be used to enhance the corresponding anisotropy parameter. Similarly, the compressive strain also reduces other magnetic properties, like magnetic moments of Vanadium, exchange parameters, and Curie temperature, which are strongly suppressed at higher compressive strains.
\section{Acknowledgments}
\label{sec:ref}
This work has been supported by the Norwegian Financial Mechanism 2014-2021 under the Polish-Norwegian Research Project NCN GRIEG “2Dtronics” no. 2019/34/H/ST3/00515.
\appendix
\section{Band Structures}

To  have a dipper insight into the impact of strain on electronic properties, we have also calculated the electronic  band structures of both strained and unstrained structures. The numerical calculations were performed  within the Generalized Gradient Approximation with Hubbard U correction (GGA+U) scheme, as well as in the Generalized Gradient Approximation with Spin-Orbit Coupling (GGA+SOC) method. 
The band structures for strain levels of $-8 \% $, $-4 \% $, $0 \% $, $+4 \% $, and $+8 \%$ are shown in Fig.8, and the corresponding band gaps are given  in Table 1. 
This figure and data in the Table 1 clearly show that the electronic structure changes remarkably under the external stain.



\begin{figure*}
    \centering
    \includegraphics[width=2\columnwidth]{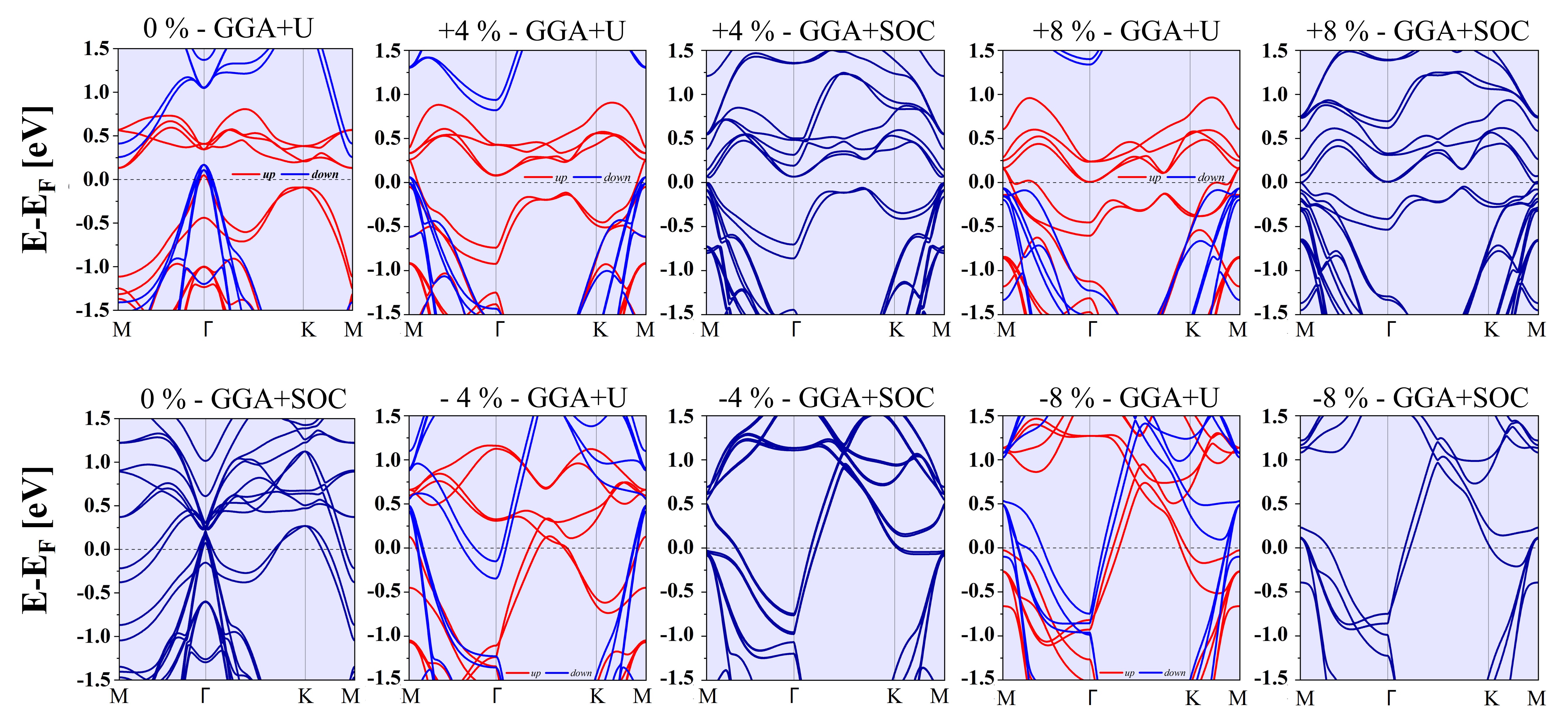}
    \caption{Spin-resolved band structures for the Bilayer T-phase of VS2 in the GGA+U and GGA +SOC approximations.}
    \label{bandstructure}
\end{figure*}

\setcounter{table}{0}
\setcounter{figure}{0}
\renewcommand{\thefigure}{\thesection.\arabic{figure}}
\renewcommand{\thetable}{\thesection.\arabic{table}}

 \bibliographystyle{elsarticle-num} 
 \bibliography{Ref.bib}
\end{document}